\documentclass[preprint,superscriptaddress,nofootinbib]{revtex4}
\usepackage{amsmath}
\usepackage{graphicx}
\usepackage{color}
\usepackage{subfigure}

\newcommand\nn{\nonumber}
\newcommand\beq{\begin{equation}}
\newcommand\eeq{\end{equation}}
\newcommand\bea{\begin{eqnarray} }
\newcommand\eea{\end{eqnarray} }

\begin{document}

\rightline{CYCU-HEP-14-02}
\bigskip
\title{Spin chains and classical strings in rotating Rindler-AdS space}
\bigskip
\author{Shou-Huang Dai}
\email{shdai.hep@gmail.com}
\affiliation{Leung Center for Cosmology and Particle Astrophysics, National Taiwan University, Taipei 106, Taiwan}

\author{Shogo Kuwakino}
\email{shogo.kuwakino@gmail.com}
\affiliation{Department of Physics and Center for High Energy Physics, Chung Yuan Christian University, Chung Li City, Taiwan}

\author{Wen-Yu Wen}
\email{steve.wen@gmail.com}
\affiliation{Leung Center for Cosmology and Particle Astrophysics, National Taiwan University, Taipei 106, Taiwan}
\affiliation{Department of Physics and Center for High Energy Physics, Chung Yuan Christian University, Chung Li City, Taiwan}

\begin{abstract}
In this paper, we study the spin chain and string excitation in the rotating Rindler-$AdS_3$ proposed in \cite{Parikh:2011aa}.   We obtain a one-parameter deformed $SL(2)$ spin chain at the fast spin limit.  Two-spin GKP-like solutions are studied at short and long string limits.  General ansatz for giant magnons and spiky strings are analyzed in detail for various $\beta$.  At last, we explore its counterpart in analytic continuation  and pp-wave limit.
\end{abstract}

\maketitle



\section{Introduction and Summary}

AdS/CFT correspondence \cite{holo1,holo2,holo3} opens a new window for us to understand the fundamental physics of the nature. This conjecture had been tested from many aspects. One of them is the duality between the integrable spin chains arising from the single trace operators in $\mathcal{N}=4$ super Yang-Mills (SYM) theory \cite{spinchainSYM} and the rotating strings in $AdS_5 \times S^5$ \cite{dualstring,Kruczenski:2003gt,Bellucci:2004qr}. This duality can be checked by the agreement between the spin chain effective action and the rotating string sigma model action \cite{Kruczenski:2003gt,Bellucci:2004qr}, and from the view point of integrability such that the Bethe equation for the spin chain and that for the classical string sigma model in $AdS_5 \times S^5$ are equivalent \cite{integ}. The spiky string solution dual to the giant magnon excitation of the spin chain are also constructed \cite{Hofman,Ryang:2006yq}.

One may wonder the possibility of generalizing the AdS/CFT correspondence to the non-AdS/non-CFT cases. A way to investigate this problem is to consider deformations on both sides and see how the correspondence works. One example in the context of the super Yang-Mills spin chain is the $\beta$-deformed $\mathcal{N}=4$ SYM and its gravity dual \cite{Lunin}. In contrast to \cite{Lunin} where a smooth deformation from $S^5$ in the original $AdS_5 \times S^5$ background is accounted for, in this paper we explore the spin chain dual solution in a gravitational background which switches to another conjugacy class of $AdS_3$ once the deformation is turned on.

One-parameter stationary vacua\footnote{
Note that the deformation parameter of our $AdS$ $\beta$-vacua has nothing to do with that of the $\beta$-deformed SYM. We just follow the convention in the literature, and remind the readers to mind the possible confusion.}
in $AdS_3$ belonging to the loxodromic conjugacy class of the Lorentz group (i.e. rota-boosts) were constructed in \cite{Parikh:2011aa} by twisting the Lorentz group in the embedding flat space. In particular, a combination of two boosts $M_{01} - \beta M_{23}$ (where $x_0$ and $x_3$ are time-like coordinates) gives rise to the rotating Rindler $AdS$ space, which is the universal cover for the BTZ black hole but with less symmetry. There is an event horizon and an ergosphere due to this rota-boost. As the deformation (or rotation) parameter $\beta$ vanishes, it reduces to the Rindler spacetime, with the acceleration identified as the inverse of the AdS radius, $a \sim L^{-1}$. On the other hand, a combination of a temporal and a spatial rotations $M_{03}-\beta M_{12}$ gives rise to the rotating global $AdS$, where some region around the center is hidden from a co-rotating observer. As $\beta$ vanishes, it reduces to the AdS in global coordinates. Although the boundary theory dual to the rotating global $AdS$ still satisfies the Virasoro algebra, the conformal symmetry in static vacuum is broken for uneven deformation in the left and the right sectors. While the computation power seems out of control on the field theory side due to lost symmetry, one is hoping that computation from gravity side is still tractable either analytically (at certain limits) or numerically. As $\beta$ vanishes, the rotating Rindler $AdS_3$ falls in the conjugacy class of hyperbolic transformations (i.e. pure boosts), while the global rotating $AdS_3$ switches to the class of elliptic transformations (i.e. pure rotations). Note that the rota-boosts cannot reduce to the pure boosts or the pure rotations by Lorentz transformations.

The goal of this paper is to study the rotating global $AdS$ vacuum by probing a classical string and observe the effect of deformation to the string excitation and dispersion. We solve the classical string solutions dual to the spin chain in the $\beta$-vacua of global rotating $AdS_3 \times S^1$ embedded in $AdS_5 \times S^5$. At certain limits, we can also obtain the analytic expressions for spin chain model and dispersion relation. Their implications on the dual field theory side, however, remains to be investigated.

The structure of this paper is outlined as follows. In Section II, we derive a deformed $SL(2)$ spin chain Hamiltonian for a fast spinning string, and some simple excitation is examined for nonzero deformation.  In Section III, we first study the dispersion relation for GKP string at different limits, and then the general solutions for giant magnons and spiky strings are analyzed in detail.  In Section IV, we study the analytically continued version of the geometry and its sin(h)-Gordon model. In Section V, the dispersion relation for a spiky string is studied in the pp-wave limit.


\section{A spin chain from AdS $\beta$-vacua}

It was shown in \cite{Kruczenski:2003gt} that in the fast spinning limit, one was able to obtain the Heisenberg spin chain by using the sigma model approach, which agrees with the one-loop calculation of anomalous dimensions in $\mathcal{N}=4$ super Yang-Mills.   Although this quantity is no longer protected by the symmetry in the deformed theory, one can still study the effect of deformation to the spin chain Hamiltonian and dynamics from gravity side.  In order to reach sensible spinning limit, we will include additional circle in the background metric.  This circular dimension is easily obtained from dimensional reduction, say from type IIB theory in ten dimensions to $AdS_3\times S^3\times T^4$.   After deforming to the rotating global $AdS_3$\cite{Parikh:2011aa} , the global metrics reads:
\begin{equation}\label{globalmetric}
ds^2 = -((1-\beta^2)\cosh^2{\rho}+\beta^2)dt^2 + 2\beta dt d\theta +d\rho^2 + ((1-\beta^2)\cosh^2{\rho}-1)d\theta^2 + d \phi^2
\end{equation}
Note that in this coordinate, $\cosh^2 \rho < \frac{1}{1-\beta^2}$ is inaccessible due to the deformation, since in this region the rotation generator $\frac{\partial}{\partial \theta}$ becomes time-like. Now we would like to show that at the fast spinning limit, a spin chain Hamiltonian can be obtained from the string worldsheet.  Without loss of generality, we apply the following embedded ansatz:
\begin{equation}
t=\kappa \tau, \quad \rho = \rho(\tau,\sigma),\quad \theta = \theta(\tau,\sigma), \quad \phi = \phi(\tau,\sigma),
\end{equation}
where $(\tau,\sigma)$ are worldsheet coordinates.  A change of coordinates
\begin{equation}
\theta \to \theta + t,\quad \phi \to \phi + (1-\beta)t, \quad \rho \to \frac{1}{2}\rho,
\end{equation}
brings the Polyakov action into\footnote{
The general form of the equations of motion and the Virasoro constraints arising from the Polyakov action is presented in the Appendix.}
\begin{eqnarray}
S = &&\frac{\sqrt{\lambda}}{4\pi}\int{d\tau d\sigma} \{ \frac{1}{4}(\dot{\rho}^2-\rho^{\prime 2})+[(1-\beta^2)\cosh^2{\frac{\rho}{2}}-1 ](\dot{\theta}^2-\theta^{\prime 2})+2[(1-\beta^2)\cosh^2 \frac{\rho}{2}\nonumber\\
&&-(1-\beta)]\kappa \dot{\theta} +(\dot{\phi}^2 -\phi^{\prime 2}) + 2(1-\beta)\kappa \dot{\phi} \},
\end{eqnarray}
where we use $X^\prime$ to denote the derivative of $X$ with respect to $\sigma$ and $\dot{X}$ with respect to $\tau$.  Then we take fast spinning limit\cite{Kruczenski:2003gt}, by sending $\kappa \to \infty$ and $\dot{X^{\mu}}\to 0$, such that $\kappa \dot{X^{\mu}}$ remains finite.  After taking the limit, the above action simplifies as
\begin{equation}\label{ads-spinchain}
S = \frac{\sqrt{\lambda}}{4\pi}\int{d\tau d\sigma} \{2[(1-\beta^2)\cosh^2{\frac{\rho}{2}}-(1-\beta)]\kappa\dot{\theta}  + 2(1-\beta)\kappa\dot{\phi} - \frac{1}{4}\rho^{\prime 2} - [(1-\beta^2)\cosh^2{\frac{\rho}{2}}-1]\theta^{\prime 2}-\phi^{\prime 2} \}.
\end{equation}
Taking account of the Virasoro constraint: $G_{\mu\nu}\dot{X^{\mu}}X^{\nu}{ }^\prime=0$, that is
\begin{equation}
[(1-\beta^2)\cosh^2{\frac{\rho}{2}}-(1-\beta)]\theta^\prime + (1-\beta)\phi^\prime = 0,
\end{equation}
one reaches the classical action of spin chain
\begin{equation}\label{eqn:Landau_Lifshitz}
S = \frac{\sqrt{\lambda}}{4\pi}\int{dt d\sigma} \{(1-\beta)[(1+\beta)\cosh{\rho}-(1-\beta)]\partial_t{\theta}  + 2(1-\beta)\partial_t{\phi} - \frac{\tilde{\lambda}}{2L^2} {\cal H} \},
\end{equation}
where we define $\tilde{\lambda}\equiv \frac{\lambda}{8\pi^2}$ and $L \equiv \frac{\kappa \sqrt{\lambda}}{ 2 \pi }$, and the spin chain Hamiltonian density reads
\begin{equation}
{\cal H}=(\partial_\sigma\rho)^2 + (1+\beta)^2\sinh^2{\rho}(\partial_\sigma \theta)^2.
\end{equation}
This will reduce to $SL(2)$ XXX spin chain at the limit $\beta \to 0$ as expected\cite{Bellucci:2004qr}.

\begin{figure}
\begin{center}
 \subfigure[{\cal E} v.s. $\omega$ for a rotating string]
  {
    \includegraphics[width=0.45\textwidth]{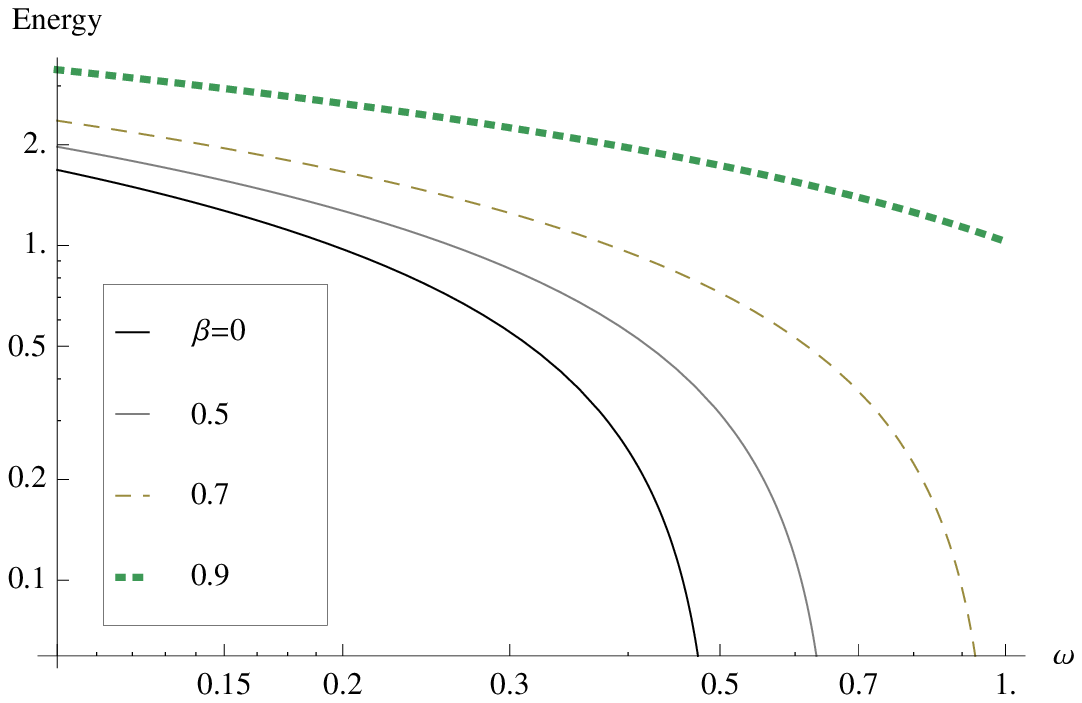}
    \label{fig:energy}
  }
  \subfigure[$S$ v.s. $\omega$ for a rotating string]
  {
     \includegraphics[width=0.45\textwidth]{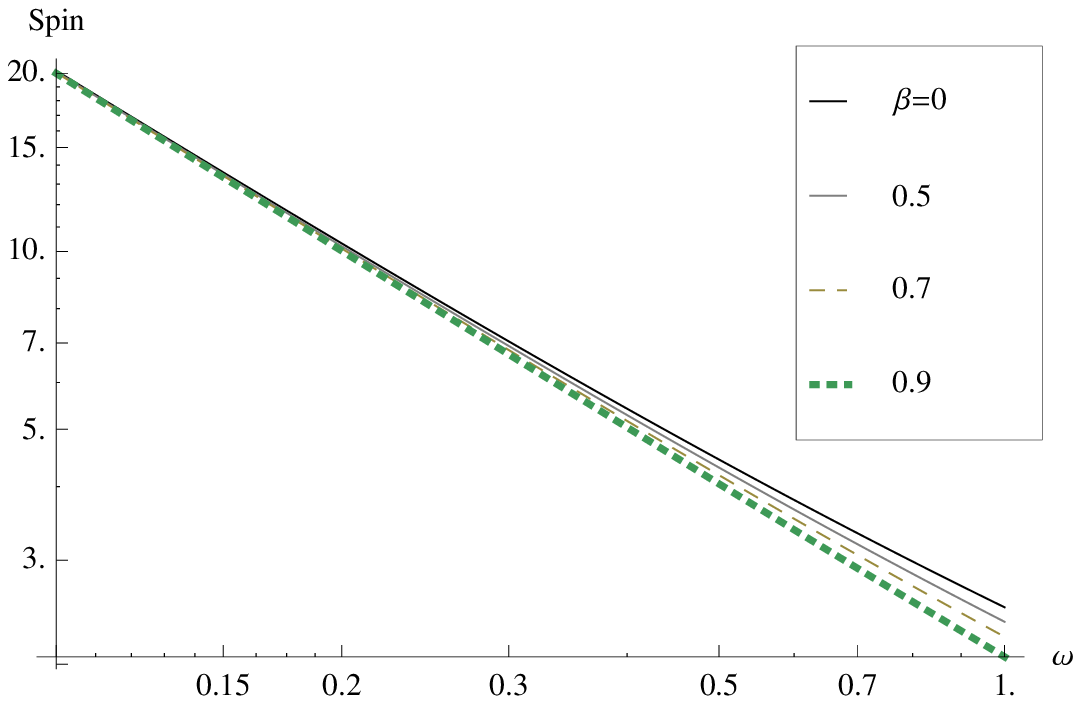}
     \label{fig:spin}
  }
\caption{(a) shows the log plot for energy of a rotating string for deformation parameter $\beta=0$(solid black), $0.5$(solid gray), $0.7$(dashed brown), $0.9$(dotted green).  (b) shows the spin of a rotating string for the same deformation parameters.
}\label{fig:rotating_string}
\end{center}
\end{figure}

To illustrate the effect of deformation, we will examine a rotated string solution against the deformed spin chain action (\ref{eqn:Landau_Lifshitz}).  First, the equations of motion are derived:
\begin{eqnarray}
(1-\beta^2)\partial_t \cosh{\rho} - \frac{\tilde{\lambda}}{L^2}(1+\beta)^2\partial_\sigma (\sinh^2{\rho} \partial_\sigma \theta)&=&0,\nonumber\\
(1-\beta^2)\sinh{\rho}\partial_t \theta + \frac{\tilde{\lambda}}{L^2}[\partial_\sigma^2 \rho - \frac{1}{2}(1+\beta)^2\sinh{2\rho} (\partial_\sigma \theta)^2]&=&0.
\end{eqnarray}
The simplest solution is obtained for $\theta = \omega t$ and $\rho=\rho(\sigma)$.  Then the equation of motion for $\rho$ can be integrated to give
\begin{equation}\label{lleom}
\partial_\sigma \rho = \pm \sqrt{a-b \cosh{\rho}}, \qquad b=\frac{2L^2}{\tilde{\lambda}}(1-\beta^2)\omega,
\end{equation}
for some constant $a$.  This solution describes that a folded string stretching between $\rho=\pm \rho_{max} = \cosh^{-1}{\frac{a}{b}}$ rotates in uniform speed at the center of $AdS_3$.  The total energy $E$ and spin $S$, defined as follows, can be written in terms of complete elliptic integral of the first kind $K(x)$ and second kind $E(x)$:
\begin{eqnarray}
&&E = \frac{\tilde{\lambda}}{4L}\int{d\sigma (\partial_\sigma \rho)^2} = -2\sqrt{2b} \frac{\tilde{\lambda}}{L} \{ E(x)-(1-x)K(x) \},\nonumber\\
&&S = \frac{L}{2}\int{d\sigma (1-\beta^2)\cosh{\rho}} = 2(1-\beta^2)L\sqrt{\frac{2}{b}} \{ 2E(x)-K(x)\},
\end{eqnarray}
where $x=\frac{b-a}{2b}$.  We plot the energy and spin against $\omega$ for several $\beta$'s in the Figure (\ref{fig:rotating_string}) and find out that deformation increases the energy but slows down the spin.

Several comments are in order:  first, the horizon censoring the $AdS$ center seems boosted away in the fast spin limit such that the folded string is able to pass through $\rho=0$.   Secondly, the equation (\ref{lleom}) implies that the apparent string tension is enhanced by a factor $(1-\beta^2)^{-1/2}$.   We recall in the earlier studies of turning on the NSNS field $B$ for spinning string in $S^3$\cite{Chen:2008vc,Lee:2008sk}, the apparent tension is reduced by a factor $(1-B^2)^{1/2}$.   This might be some kind of electric-magnetic duality or strong-weak duality between the metric component $G_{t\theta}$ and its analytic continued counterpart $G^{\prime}_{\varphi_1\varphi_2}$, which acts as a nontrivial $B_{\varphi_1\varphi_2}$ on the string worldsheet\footnote{For the analytic continuation of the metric (\ref{globalmetric}), see the change of coordinates  (\ref{analytic_continue}) in the section IV.}.

\section{Two-spin strings}
There are three $U(1)$'s found in the isometry $SO(2,2)\times SO(2)$ for the target space (\ref{globalmetric}).   While a $U(1)$ is identified to the global time generator, one can still turn on maximal two spins charged under the remaining $U(1)^2$.  In the following sections, we will first study the spinning closed string solution to understand its leading Regge trajectory behavior at short and long string limits.  Then we will construct general ansatz for giant magnons and spiky strings and obtain their dispersion relations.

\subsection*{GKP solution}

First we consider a two-spin classical solution in $\beta$-deformed $ AdS_3 \times S^1$ background\footnote{In \cite{Gubser:2002tv}, a spinning string in $ AdS_3$ space was originally discussed. Also a two-spin solution in $ AdS_3 \times S^1$ space was discussed in \cite{Frolov:2002av}.}.  We apply the following ansatz to the metric (\ref{globalmetric}):
\begin{equation}
t = \kappa \tau, \ \rho = \rho ( \sigma ), \ \theta =  \omega \tau, \ \phi =  \chi \tau,
\end{equation}
where $\kappa, \omega$ and $\chi$ are integers. In the conformal gauge, the Polyakov action leads to
\begin{eqnarray}
S &=& \frac{\sqrt{\lambda}}{4\pi }
\int d\tau d\sigma
\left\{ -  [ (1-\beta^2) \cosh^2 \rho + \beta^2 ]  \kappa^2
+2\beta \kappa \omega
- (  \partial_\sigma \rho )^2 \right. \nn \\
&& \hspace{2.4cm} \left. +  [ (1-\beta^2) \sinh^2 \rho - \beta^2 ]\, \omega^2 + \chi^2
\right\}.
\end{eqnarray}
From the Virasoro constraint, we get the following equation
\begin{equation}
\frac{d \rho}{d \sigma} =
\pm  \sqrt{  \{ ( \kappa - \beta \omega )^2 - \chi^2 \} \cosh^2 \rho -  \{ ( \omega - \beta \kappa )^2 - \chi^2 \} \sinh^2 \rho   }.
\label{drhodsigma}
\end{equation}
We will present the solution later. For now, by integrating \eqref{drhodsigma}, we obtain the following relation
\begin{eqnarray}
2 \pi &=& 4 \int^{\rho_{{\rm max}}}_0 \frac{   d \rho }{ \sqrt{  \{ ( \kappa - \beta \omega )^2 - \chi^2 \} \cosh^2 \rho -  \{ ( \omega - \beta \kappa )^2 - \chi^2 \} \sinh^2 \rho   } } \nonumber \\
&=&  \frac{4}{ \sqrt{ ( \kappa - \beta \omega )^2 - \chi^2 } }
\frac{1}{ \sqrt{ 1+\eta } }
K ( \frac{1}{\sqrt{1+\eta}} ),
\label{pib}
\end{eqnarray}
here $\rho_{{\rm max}}$ and $\eta$ are defined as
\begin{equation}
\coth^2 \rho_{{\rm max}} =  \frac{ ( \omega - \beta \kappa )^2 - \chi^2 }{ ( \kappa - \beta \omega )^2 - \chi^2 } = 1 +\eta.
\label{rhomax}
\end{equation}

We calculate three conserved quantities of this system, the energy $E$ and two angular momenta $S$ and $J$ which are associated to the coordinates $\theta$ and $\phi$ respectively
\begin{eqnarray}
E &=& - P_t
= \frac{\sqrt{\lambda}}{2 \pi } \int_0^{2\pi } \left\{  \kappa \left( (1-\beta^2) \cosh^2 \rho + \beta^2 \right) - \beta \omega \right\} d \sigma,  \\
S &=& P_{\theta} = \frac{\sqrt{\lambda}}{2 \pi } \int_{0}^{2\pi }
\left\{   \omega ( (1-\beta^2) \sinh^2 \rho - \beta^2 )  + \beta \kappa \right\} d \sigma, \\
J &=& P_{\phi} = \sqrt{\lambda} \chi.
\label{SpinJ}
\end{eqnarray}
Using \eqref{drhodsigma}, the energy $E$ and the spin $S$ can be evaluated in terms of the elliptic integral as
\begin{eqnarray}
E &=&   \frac{2\sqrt{\lambda}}{ \pi  }
\frac{ 1 }{  \sqrt{ ( \kappa - \beta \omega )^2 - \chi^2 } }
\bigg(
\kappa (1-\beta^2)
\frac{ \sqrt{ 1+\eta } }{ \eta }
 E ( \frac{1}{\sqrt{1+ \eta}} )
- \frac{ \beta( \omega - \beta \kappa  ) }{ \sqrt{ 1+\eta } } K ( \frac{1}{\sqrt{1+ \eta}} )
\bigg),
\label{energy}
\\
S &=&   \frac{2\sqrt{\lambda}}{ \pi  }
\frac{ 1 }{  \sqrt{ ( \kappa - \beta \omega )^2 - \chi^2 } }
\bigg(
\omega (1-\beta^2)
\frac{ \sqrt{1+\eta} }{\eta} E ( \frac{1}{\sqrt{1+ \eta}} )
- \frac{\omega - \beta \kappa}{\sqrt{1+\eta}}
K (  \frac{1}{\sqrt{1+ \eta}} )
\bigg).
\label{spin}
\end{eqnarray}
In the following we consider the short string limit $\rho_{{\rm max}} \ll 1$ and the long string limit $\rho_{{\rm max}} \gg 1$, and evaluate a relation between the energy $E$ and the spins $S$ and $J$ for each case.

\subsection{Short string limit}

First we consider the short string limit $\rho_{{\rm max}} \ll 1$. From \eqref{rhomax}, this corresponds to the limit $\eta \gg 1$.
From \eqref{pib} and \eqref{rhomax}, by taking the limit, we get the relations
\begin{eqnarray}
(\kappa -\beta \omega)^2 &\sim& \chi^2 +  \frac{1}{\eta}, \\
(\omega - \beta \kappa )^2 &\sim& \chi^2 + 1 + \frac{1}{\eta}.
\end{eqnarray}
Also, by a suitable combination of \eqref{energy} and \eqref{spin}, we obtain a simple expansion relation as
\begin{equation}
S - \beta E
\sim
\frac{ \sqrt{\lambda} ( 1- \beta^2) }{ 2 }
\frac{ \sqrt{\chi^2 + 1} }{\eta}.
\label{SBeta}
\end{equation}
It is natural to identify $E-\beta S$ and $S- \beta E$ as the {\sl twisted} energy and spin in the rotating global AdS background, because they correspond to generators $\partial_t-\beta\partial_\theta$ and $\partial_\theta - \beta \partial_t$ respectively.  By substituting this, we obtain the relation of the twisted energy and the twisted spin for any value of $\chi$ as
\begin{eqnarray}
E - \beta S
&\sim & \sqrt{\lambda} (1- \beta^2 ) \sqrt{ \chi^2 + \frac{2}{\sqrt{\lambda}}  \frac{S-\beta E}{(1-\beta^2) \sqrt{ \chi^2 +1 }  } } \nn \\
&& \quad + ( S- \beta E)
\frac{ \sqrt{\chi^2 + \frac{2}{\sqrt{\lambda}}  \frac{S-\beta E}{(1-\beta^2) \sqrt{ \chi^2 +1 }  } }  }
{ \sqrt{ \chi^2 + 1+ \frac{2}{\sqrt{\lambda}}  \frac{S-\beta E}{(1-\beta^2) \sqrt{ \chi^2 +1 }  } } }.
\label{BetaShort}
\end{eqnarray}
In order to describe this relation in terms of the spin $J$, let us first consider the limit of $\chi \ll 1$. At this limit, \eqref{SBeta} leads to
\begin{equation}
S - \beta E
\sim
\frac{  \sqrt{\lambda} ( 1- \beta^2) }{ 2 }
\frac{ 1 }{\eta},
\label{SBetaShort1}
\end{equation}
and \eqref{BetaShort} is simplified to
\begin{equation}
E- \beta S \sim  \sqrt{\lambda} ( 1- \beta^2 ) \sqrt{  \chi^2 + \frac{2}{ \sqrt{\lambda}} \frac{S- \beta E}{(1- \beta^2 )}  }.
\end{equation}
Then, using \eqref{SpinJ}, we obtain the relation
\begin{equation}
\left( \frac{E-\beta S}{1-\beta^2} \right)^2
\sim J^2 + 2 \sqrt{\lambda} \frac{S- \beta E}{1-\beta^2} + \cdots
.
\end{equation}
Next let us see another limit $\chi \gg 1$. At this limit, \eqref{SBeta} leads to
\begin{equation}
S - \beta E
\sim
\frac{  \sqrt{\lambda} ( 1- \beta^2) }{ 2 }
\frac{ \chi }{\eta},
\end{equation}
and \eqref{BetaShort} becomes
\begin{equation}
E - \beta S \sim \sqrt{\lambda} ( 1- \beta^2 ) \chi + S-\beta E + \frac{S-\beta E}{\chi^2},
\end{equation}
or, using \eqref{SpinJ}, this corresponds to the relation
\begin{equation}
E - \beta S
\sim (1- \beta^2 ) J + ( S - \beta E ) - \frac{1}{2}  \frac{\lambda}{J^2}  ( S- \beta E ) + \cdots
.
\end{equation}
In the case of $\beta = 0$, this result reduces to (3.24) in \cite{Frolov:2002av} as expected.

\subsection{Long string limit}

Next we consider the case of the long string limit $\rho_{{\rm max}} \gg 1$, i.e. $\eta \ll 1$. From \eqref{pib} and \eqref{rhomax}, by taking the limit, we get the relations
\begin{eqnarray}
( \kappa - \beta \omega )^2 &\sim& \chi^2 + \frac{1}{\pi^2} \ln^2 ( \frac{1}{\eta} ), \\
( \omega - \beta \kappa )^2 &\sim& \chi^2 + ( 1 +\eta ) \frac{1}{\pi^2} \ln^2 ( \frac{1}{\eta} ).
\end{eqnarray}
Also, from the expansion of the spin $S$ in \eqref{spin}, we obtain
\begin{equation}
S \sim
\frac{2 (1+ \beta ) \sqrt{\lambda} }{\pi \eta }
\frac{\sqrt{\chi^2 + \frac{1}{\pi^2} \ln^2 \frac{1}{\eta} }}{ \sqrt{ \frac{1}{\pi^2} \ln^2 \frac{1}{\eta} }  }.
\label{SpinL}
\end{equation}
We find that the spin is large $S \gg 1$ for any value of $\chi$. Since it is not easy to spot a dispersion relation in this complicated expression, we consider several limits of $\chi$ in the following.
First let us consider the case for $ \chi \ll \ln \frac{1}{\eta}$. For the limit \eqref{SpinL} leads to
\begin{equation}
S  \sim \frac{2 \sqrt{\lambda} }{ \pi  } \frac{(1+  \beta )}{\eta},
\end{equation}
and using \eqref{SpinJ} we obtain the following relation
\begin{equation}
E
\sim
S + ( 1- \beta )\frac{ \sqrt{\lambda} }{ \pi  }
\ln  \frac{ S}{ \sqrt{\lambda} ( 1+\beta )  }
+ \frac{(1- \beta) \pi }{2 \sqrt{\lambda} }
\frac{J^2}{ \ln  \frac{ S}{ \sqrt{\lambda} ( 1+\beta )  }  }
.
\end{equation}
In the case of the opposite limit $ \chi \gg \ln \frac{1}{\eta}$, \eqref{SpinL} leads to
\begin{equation}
S \sim
2 \sqrt{\lambda} ( 1+ \beta ) \frac{\chi}{\eta}.
\label{SpinL2}
\end{equation}
From this equation we find that $S \gg \chi$. Comparing an expansion of the energy \eqref{energy} with \eqref{SpinL2}, and using \eqref{SpinJ}, we obtain the relation
\begin{equation}
E \sim S +
(1- \beta ) J
+ (1 - \beta )
\frac{ \lambda}{2 \pi^2 J}
\left(\ln  \frac{S}{ ( 1- \beta ) J}\right)^2 ,
\end{equation}
which also leads to (3.32) of \cite{Frolov:2002av} when $\beta =0$.


\subsection*{Giant magnon/Spiky solution}
\setcounter{subsection}{0}
\subsection{The solution}
To describe the 2-spin giant magnon/spiky string solutions, we take the ansatz following \cite{Ryang:2006yq}:
\beq \label{ansatz}
t=\tau+h_1(y), \quad \rho = \rho(y),\quad \theta = \omega [\tau + h_2(y)],\quad \phi= \Omega\, \tau,
\eeq
where the worldsheet coordinates are changed from $(\tau,\sigma)$ to $(\tau,y)$, with $y=\sigma-v \tau$ and $0<v<1$.  In the following, we will set $\Omega =1$ in our analysis for convenience, and $0< \omega <1$. Then the equations of motion (\ref{eomt}) and (\ref{eomtheta}) are rewritten into the differential equations for $h_1(y), h_2(y)$,\footnote{
We set $\frac{\sqrt{\lambda}}{2 \pi}=1$ in the following numerical analysis of the giant magnon/spiky sting solutions.}
\bea
\{ v g_{tt} + (1-v^2) g_{tt} h_1' + \beta \omega (1-v^2) h_2'\}'&=&0, \label{eomh}\\
\{v \,\omega\, g_{\theta\theta} + (1-v^2)\, \omega\, g_{\theta\theta}\, h_2' + \beta (1-v^2) h_1'\}'&=&0, \label{eomhh}
\eea
where now ${}'$ stands for $d/dy$. These equations further reduce to
\begin{eqnarray}
h_1' &=& -\frac{1}{1-v^2}
\frac{\beta \omega c_2 + (\beta \omega v + c_1) g_{\theta\theta}-v g_{tt} g_{\theta\theta}}{(1-\beta^2)^2 \sinh^2\rho \cosh^2 \rho}, \label{deh1}\\
h_2' &=& -\frac{1}{1-v^2}
\frac{-\beta c_1 + (\beta v -\omega c_2) g_{tt} - \omega v g_{tt} g_{\theta\theta}}{\omega(1-\beta^2)^2 \sinh^2\rho \cosh^2 \rho},\label{deh2}
\end{eqnarray}
where $c_1, c_2$ are two integration constants arising from integrating $(\ref{eomh})$, $(\ref{eomhh})$. As $\beta=0$, ($\ref{deh1}$) and ($\ref{deh2}$) reduce to the equations presented in \cite{Ryang:2006yq}.  The equation of motion for $\rho$ in (\ref{eomrho}) becomes
\bea
&& \rho'' = \frac{1-\beta^2}{(1-v^2)^2} \cosh \rho \, \sinh \rho \left\{
\left[ 1-\left(
\frac{\beta \omega c_2 -(c_1 + \beta \omega v + v)
\beta^2}{(1-\beta^2)^2 \sinh^2 \rho\, \cosh^2 \rho}
+\frac{c_1 + \beta \omega v}{(1-\beta^2) \cosh^2 \rho}
\right)^2
\right] \right. \nn \\
&& \hspace{2.5cm} \left.
- \omega^2 \left[ 1- \left(
\frac{- \beta c_1 /\omega + (c_2 - \beta v/\omega -v)
\beta^2}{(1-\beta^2)^2 \sinh^2 \rho\, \cosh^2 \rho}
+\frac{c_2 - \beta v / \omega}{(1-\beta^2) \sinh^2 \rho}
\right)^2
\right]
\right\}.
\label{eomrho2}
\eea

With the string profile ansatz (\ref{ansatz}), the Virasoro constraints are given by
\bea
g_{tt} (1-v h_1') h_1' + \beta \omega (h_1'+h_2'-2 v h_1' h_2')
+ \omega^2 g_{\theta\theta}  (1-vh_2')h_2' -v (\rho')^2 &=& 0, \\
g_{tt} [1-2v h_1' +(1+v^2) (h_1')^2] + 2\beta\omega [1-v(h_1'+h_2') +(1+v^2)h_1' h_2'] &&\nn \\
+\omega^2  g_{\theta\theta} [1-2v h_2' +(1+v^2) (h_2')^2]+(1+v^2) (\rho')^2 +1 &=&0.
\eea
Eliminating $(\rho')^2$ by equating the LHS of these two equations and substituting in the $h_1', h_2'$ expressions from (\ref{deh1}) and (\ref{deh2}), one obtains the following relation
\beq \label{c1c2rel}
c_1-\omega^2 c_2 +2 \beta \omega v +v =0
\eeq
for the two integration constants.

In order to proceed to obtain the explicit string solutions, we need to assign specific values to $c_1, c_2$. In this paper we set\footnote{
As being demonstrated later in this paper, the choice of $c_1,c_2$ in (\ref{c1c2}) gives rise to consistent $\beta=0$ reduction. One may choose other $c_1,c_2$, for example
\beq \label{altc1c2}
c_1 = -v - \beta \omega v, \qquad c_2 = \beta v/\omega.
\eeq
But when $\beta$ is taken to zero, such a choice does not yield the same condition to distinguish the hanging string and the spiky string profiles as the $\beta$-free case in \cite{Ryang:2006yq}, despite that $\beta=0$ in (\ref{c1c2}) and (\ref{altc1c2}) reduce to the $c_1, c_2$ choice of \cite{Ryang:2006yq} .}
\beq \label{c1c2}
c_1 = -\beta \omega v - \frac{v}{1-\beta \omega},\qquad
c_2 = -\frac{\beta^2 v}{1-\beta \omega}.
\eeq
This choice yields a constraint
\beq \label{vconstraint}
v^2 < 1 - \beta \omega
\eeq
by requiring forward propagation of the strings,
\beq
\frac{dt}{d\tau} = \frac{1}{1-v^2} \left\{
1- \frac{v^2}{(1-\beta \omega)(1-\beta^2) \cosh^2 \rho}\right\}>0.
\eeq
Note that the constraint (\ref{vconstraint}) is regarded as a natural $\beta$-deformation from the original $v^2 < 1$. Moreover, (\ref{c1c2rel}) and (\ref{c1c2}) reduce to the corresponding results in \cite{Ryang:2006yq} at $\beta = 0$.

With the given $c_1, c_2$, $h_1', h_2'$ become
\bea
h_1 (y)' &=& -\frac{v}{1-v^2} \left\{
1-\frac{1}{(1-\beta \omega)(1-\beta^2) \cosh^2 \rho}
\right\}, \\
h_2 (y)' &=& -\frac{v}{1-v^2} \left\{
1-\frac{\beta/\omega}{(1-\beta \omega)(1-\beta^2) \cosh^2 \rho}
\right\}.
\eea
Substituting these expressions into the Virasoro constraints, one obtains the differential equation for $\rho (y)$:
\beq \label{derho}
\rho'{}^2 = \frac{1}{(1-v^2)^2} \left\{
(1-\omega^2) (1-\beta^2) \cosh^2 \rho + \frac{v^2}{(1-\beta \omega)^2 \cosh^2 \rho} + (\beta - \omega)^2 - (1+v^2)
\right\}.
\eeq
It is straightforward to check that $\rho'{}^2$ is indeed an integral of $\rho''$ according to (\ref{eomrho2}) and (\ref{derho}).

The next step is to solve $\rho (y)$ profile. (\ref{derho}) can be rewritten into
\bea
\rho(y){}' &=& \pm \frac{1}{(1-v^2)\cosh \rho} \sqrt{f(\rho)}, \label{derho2}\\
f(\rho)&=&(1-\omega^2) (1-\beta^2) \cosh^4 \rho \,+\, [(\beta - \omega)^2 - (1+v^2)] \cosh^2 \rho \,+\, \frac{v^2}{(1-\beta \omega)^2}, \nn
\eea
in which one finds $\rho' \to \pm \infty$ as $\rho \to \infty$. In principle, $\rho (y)$ is solved by integrating $dy = \pm (1-v^2) (\cosh \rho) d\rho/\sqrt{f(\rho)}$ over some appropriate range on both sides, but the form of $f(\rho)$ leaves $\rho (y)$ no analytic solution. However, it can be solved numerically with the same method.

\begin{figure}
\begin{center}
 \subfigure[Hanging string $\rho(y)$ solutions]
  {
    \includegraphics[width=0.4\textwidth]{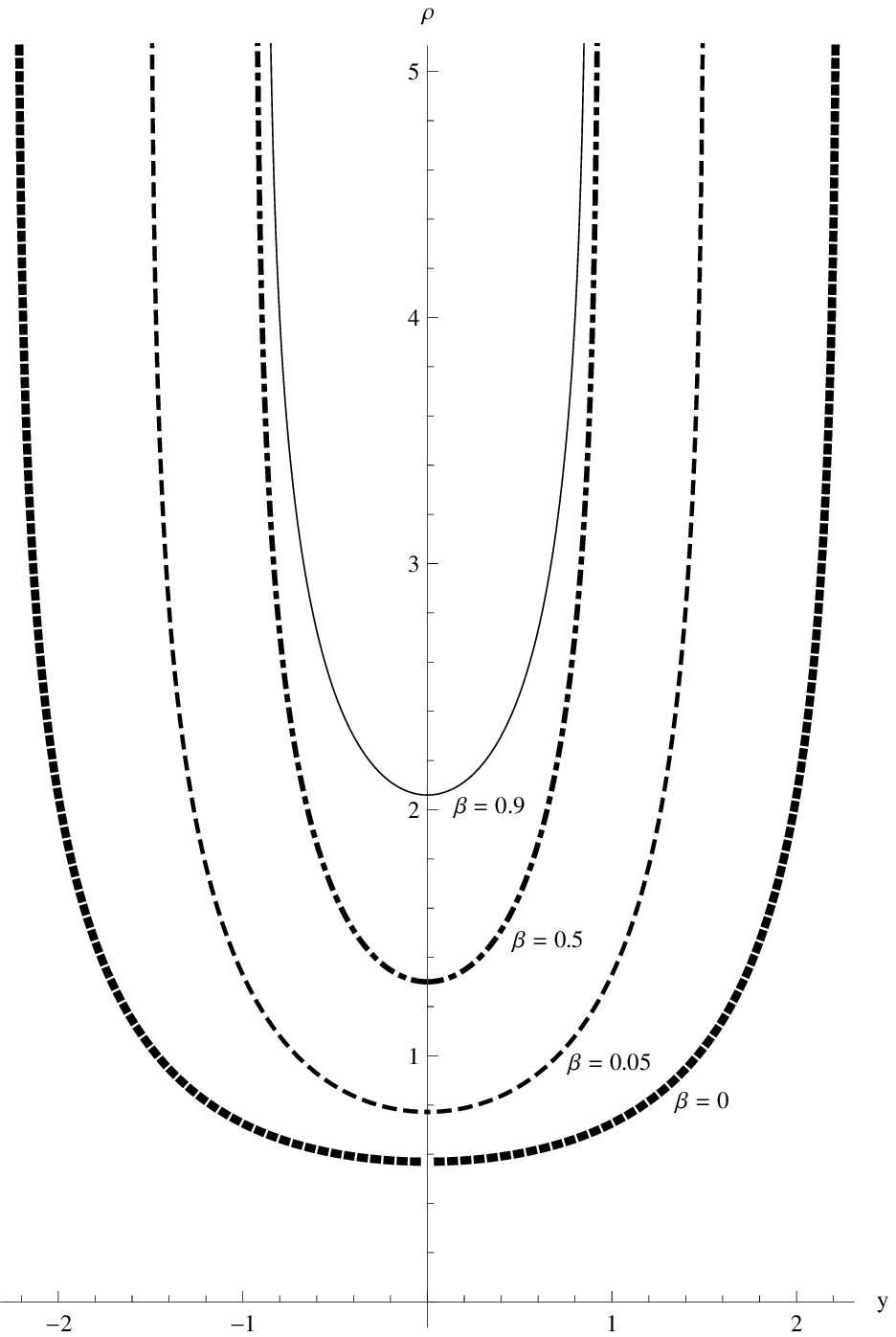}
    \label{fig:hang}
  }
  \subfigure[Bulging string $\rho(y)$ solutions at various $\beta$]
  {
     \includegraphics[width=0.55\textwidth]{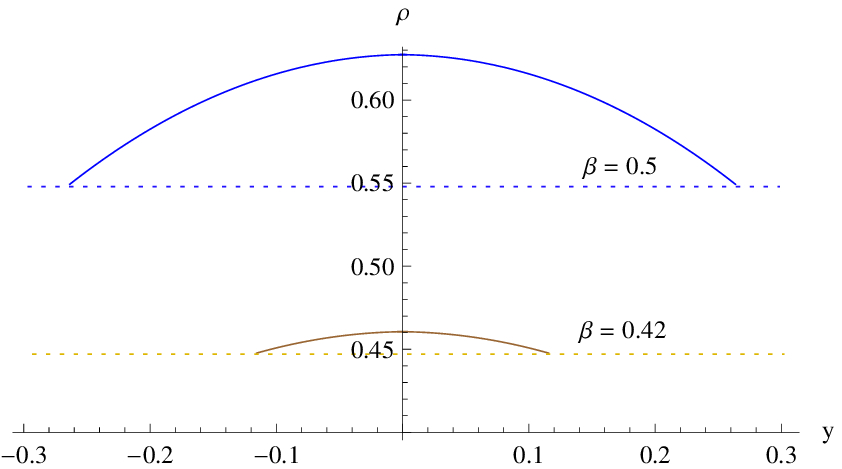}
     \label{fig:bulging}
  }
\caption{(a) shows the hanging string $\rho(y)$ profile at various $\beta$ values, with $\omega =0.8, v=0.7$. As $\beta$ increases, the width in $y$ decreases; (b) is the bulging string $\rho(y)$ at various $\beta$, with $\omega = 0.7, v = 0.8$. The dotted lines are the coordinate origins $\rho=\rho_0$ of the AdS $\beta$-vacua in eqn. (\ref{globalmetric}), with the corresponding $\beta$ values, where the strings end. $\rho_0$ increases with $\beta$.
}
\label{fig:profiny}
\end{center}
\end{figure}

\begin{figure}
\begin{center}
\includegraphics[width=0.5 \textwidth]{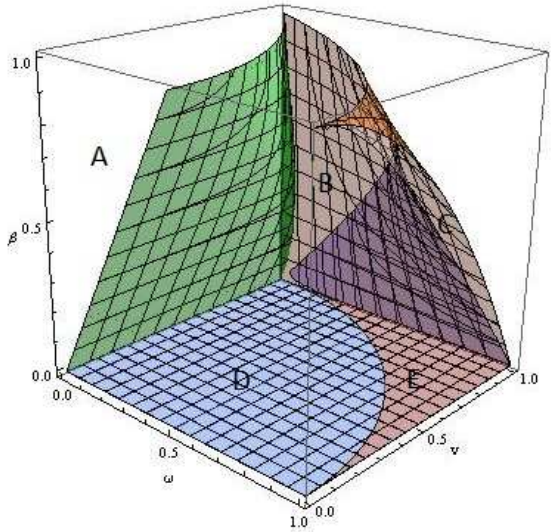}
\caption{This figure shows the parameter ranges of ($\beta$, $\omega$, v) corresponding to each of the three types of the string solutions. The green surface depicts $\rho_+ = \rho_0$, the brown surface $1-\beta \omega -v^2 = 0$, while the purple surface $\rho_- = \rho_0$. For $\beta \neq 0$, the spiky strings fall in the region A, to the left of the green surface, where $\rho_+ < \rho_0$. The hanging strings fall in the region B, in between the green and the brown surfaces, where $\rho_+ > \rho_0$ subject to the constraint $1-\beta \omega -v^2 > 0$. There is a small region C in between the purple and the brown surfaces, confined by $\rho_- > \rho_0$ and $1-\beta \omega -v^2 > 0$, corresponding to the bulging strings. At $\beta=0$ (the bottom plane), region D (the blue plane) depicts $1-\omega^2-v^2>0$ and corresponds to the spiky string solutions, while the region E (the magenta plane) has  $1-\omega^2-v^2<0$, giving rise to the hanging strings. Note that the spiky string (giant magnon) regions for vanishing and non-vanishing $\beta$'s (i.e. regions D and A respectively) are disjoint.}
\label{prange}
\end{center}
\end{figure}

$f(\rho)$ is a quadratic function in $\cosh^2 \rho$, and analysis shows that within $\beta, \omega, v \in (0,1)$, there always exist two real roots $\cosh^2 \rho_+,\, \cosh^2 \rho_-$ (where $\rho_+ > \rho_-$) for $f(\rho)$:
\begin{equation}\label{rhopm}
  \cosh^2 \rho_{\pm} = \frac{(1+v^2)-(\beta-\omega)^2 \pm \sqrt{[(1+v^2)-(\beta-\omega)^2]^2 -\frac{4 v^2 (1-\omega^2)(1-\beta^2)}{(1-\beta \omega)^2}}}{2(1-\omega^2)(1-\beta^2)}.
\end{equation}
The string solutions only exist for $f(y) \geq 0$, i.e. $\rho\geq \rho_+$ and $\rho \leq \rho_-$. On the other hand, the radial coordinate is physical for $\rho \geq \rho_0$, where $\cosh^2 \rho_0 = \frac{1}{1-\beta^2}$. Therefore the rotating string solutions can be classified by comparing $\rho_+$ and $\rho_0$:

(1) $\rho_+ > \rho_0$. This type of solution has two subclasses: (a) $\rho_- \leq \rho_0$. The strings can only extend between $\rho=\rho_+$ and $\rho=\infty$, but not for $\rho_0 \leq \rho < \rho_+$. This corresponds to the hanging strings displayed in Fig. \ref{fig:hang}. (b) $\rho_- > \rho_0$. The string can either be a hanging one like that in Fig. \ref{fig:hang}, or a bulging string confined between $\rho_0 \leq \rho \leq \rho_-$ in Fig. \ref{fig:bulging}. Note that the bulging string solution does not allow $\beta=0$ reduction, and can exist only for some limited range of $(\beta, \omega, v)$. The parameter region for this type of solution is constrained by $\rho_- > \rho_0$ and $1 - \beta \omega -v^2 > 0$ in (\ref{vconstraint}). (See Fig. \ref{prange} for numerical results.)

In this subclass, $\rho(y)'=0$ at $\rho=\rho_+$ or $\rho=\rho_-$ corresponds either to the bottom points of the hanging strings, or to the tips of the bulging stings respectively.

\begin{figure}
\begin{center}
\includegraphics[width=0.55 \textwidth]{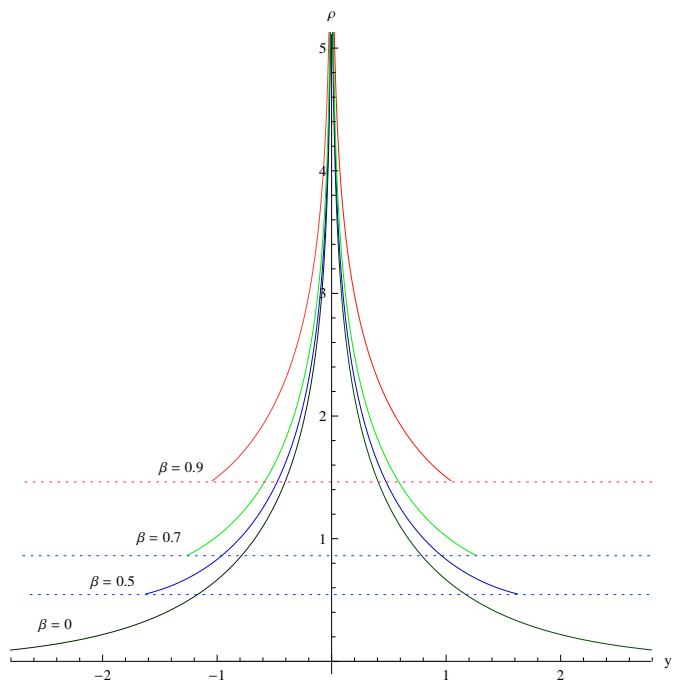}
\caption{
This figure shows the spiky string solutions $\rho(y)$ at various $\beta$, with $\omega = 0.2, v=0.4$. The dotted lines are the coordinate origins $\rho=\rho_0$ of the AdS $\beta$-vacua for each $\beta$.}
\label{fig:spiky}
\end{center}
\end{figure}

(2) $\rho_+ \leq \rho_0$. The string extends all the way from $\rho_0$ to the asymptotic infinity, and it is a spiky solution depicted in Fig. \ref{fig:spiky}.

One finds that the range of $y$ for the spiky string becomes finite due to the $\beta$-deformation of the AdS vacua. In fact, the $y$ range of all three types of solutions depends on the value of $\beta$. It is revealed in Figures \ref{fig:hang} and \ref{fig:spiky} that, for the hanging and the spiky strings, as $\beta$ increases, the width of $\rho(y)$ decreases, which implies that at constant worldsheet time $\tau$, the range of $\sigma$ shrinks as $\beta$ increases. However, since the worldsheet coordinates still have a remaining Weyl symmetry in the Polyakov action, one can rescale $\sigma$ according to $\beta$ to get rid of this issue. As for the bulging stings in Fig. \ref{fig:bulging}, the length of the string segment increases with $\beta$.


The parameter regions of $(\beta, \omega, v)\in (0,1)$ for all three types of the string solutions are shown in Fig. \ref{prange}. The boundary separating the spiky and the hanging profiles are obtained comparing $\rho_+$ and $\rho_0$, and as $\beta$ is infinitely small, it cannot be obtained as a smooth deformation from the profile distinguishing condition at $\beta=0$ derived from (\ref{derho2}).
The transition from vanishing to non-vanishing $\beta$ is discontinuous in this aspect. At $\beta=0$, the two roots in for $f(\rho)$ becomes $\cosh^2 \rho_1=1$ and $\cosh^2 \rho_2=\frac{v^2}{1-\omega^2}$, while the origin is at $\rho_0=0=\rho_1$. Here $\rho_2$ can be greater or smaller than $\rho_1$. If $\rho_2 > 1$, i.e. $\frac{v^2}{1-\omega^2}>1$, it is a hanging string; otherwise the solution is spiky (for $\frac{v^2}{1-\omega^2}\leq1$), as predicted by \cite{Ryang:2006yq}. This can be seen in Fig. \ref{prange} that the region A (the spiky strings at $\beta \neq 0$) and region C (the spiky strings at $\beta = 0$) are disjoint. This result implies that the string profile classification condition is different when the background belongs to different conjugacy classes, and they may not deform smoothly to each other, as the global rotating AdS space ($\beta \neq 0$) is in the loxodromic transformation class, while the global AdS ($\beta=0$) is in the elliptic transformation one.

\begin{figure}
\begin{center}
 \subfigure[ ]
  {
    \includegraphics[width=0.4\textwidth]{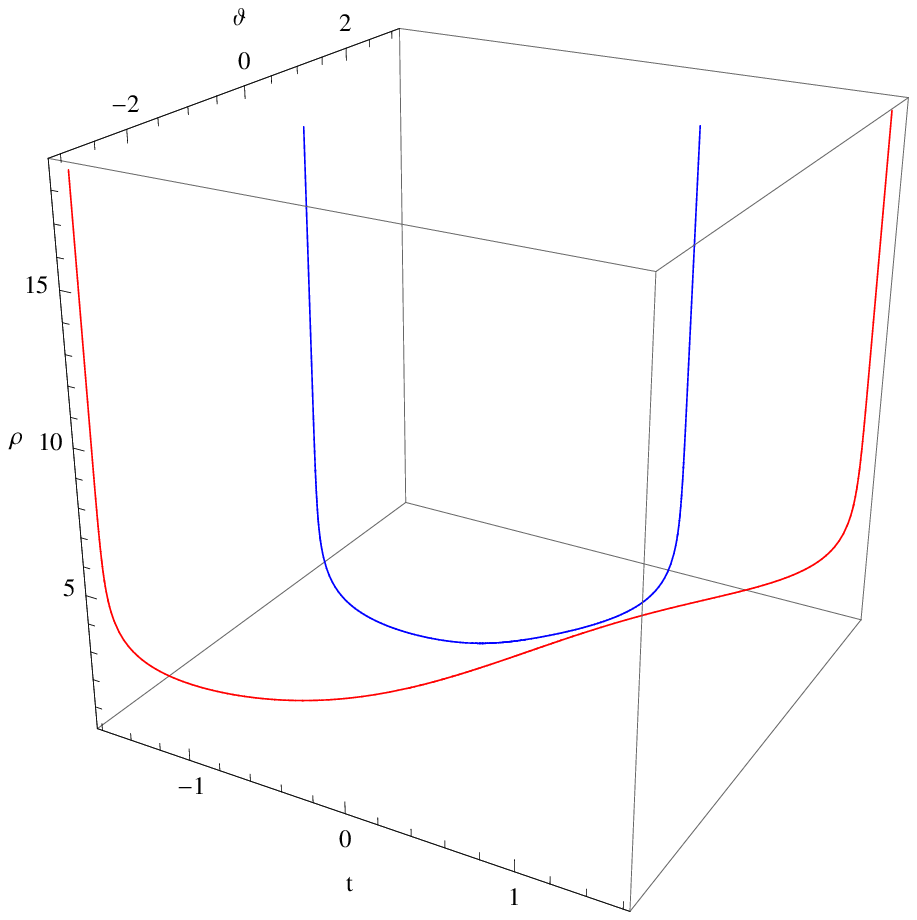}
    \label{fig:3dhang}
  }\qquad
  \subfigure[ ]
  {
     \includegraphics[width=0.4\textwidth]{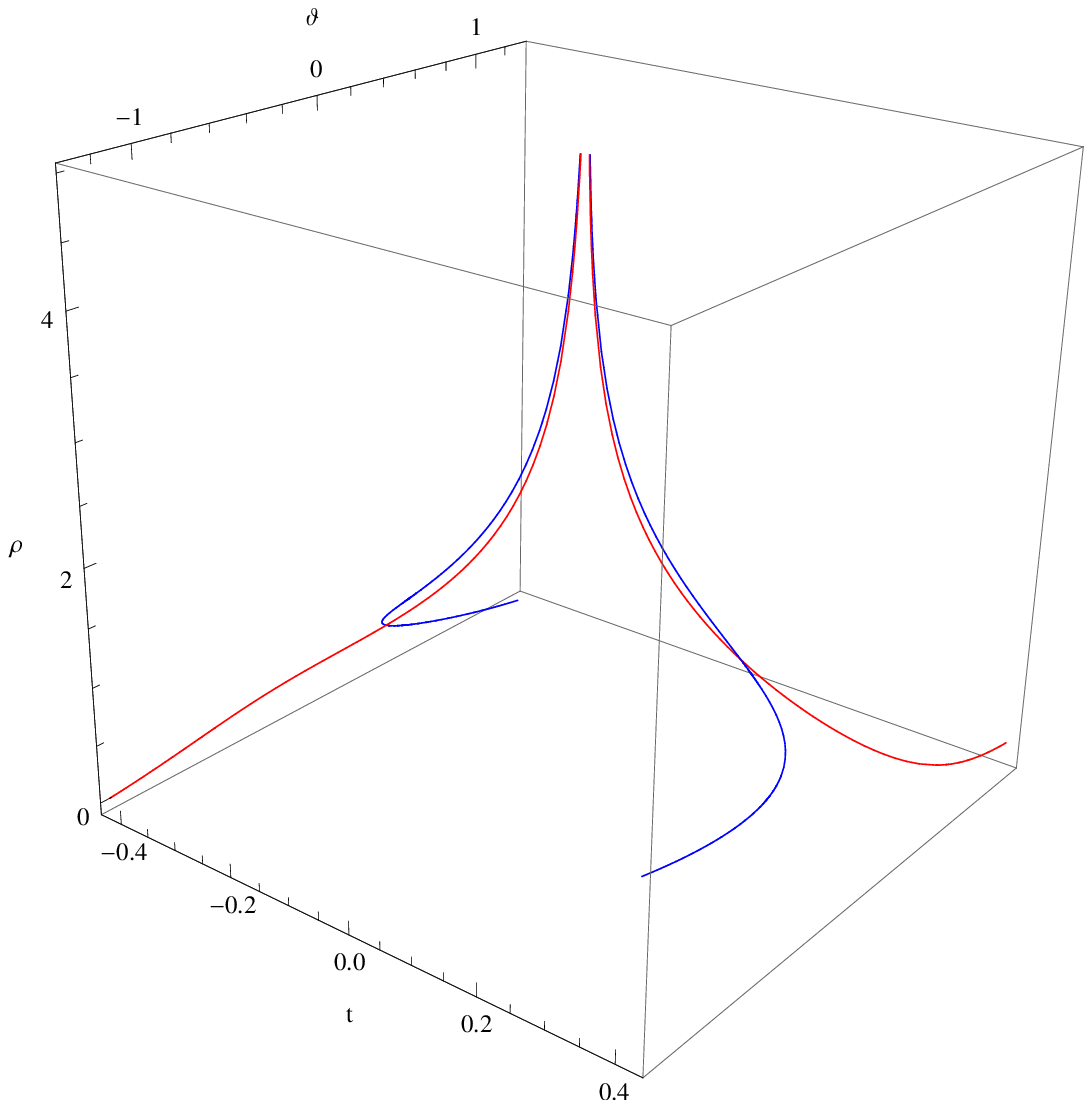}
     \label{fig:3dspiky}
  }
\caption{Rotating string profiles in the target space ($t$, $\rho$, $\theta$) with $\beta=0$ (red) and $\beta=0.5$ (blue) at $\tau=0$: (a) hanging strings, with $\omega =0.8,\, v=0.7$; (b) spiky strings, with $\omega =0.2,\, v=0.4$.
}
\label{fig:3d}
\end{center}
\end{figure}

Comparison of the snapshots (at constant worldsheet time $\tau=0$) of the spiky and the hanging strings in the $AdS_3$ $\beta$-vacua target space for $\beta=0$ and $\beta=0.5$ are shown in Fig. \ref{fig:3d}.

\subsection{Dispersion relation}

The rotating strings in the $\beta$-vacua in $AdS_3 \times S^1$ carry the following energy $E$ and spins $S, J$ which associate with $\theta, \phi$ respectively:
\bea
E &=& \frac{\sqrt{\lambda}}{2 \pi (1-v^2)} \int d\sigma \left\{
      (1-\beta^2) \cosh^2 \rho + \beta^2 -\beta \omega - \frac{v^2}{1-\beta\omega}
\right\},\label{eqE} \\
S &=& \frac{\sqrt{\lambda}}{2\pi (1-v^2)} \int d\sigma \left\{
     \omega (1-\beta^2) \sinh^2 \rho - \omega \beta^2 + \beta - \frac{v^2 \beta}{1-\beta \omega}
\right\},\label{eqS} \\
J &=& \frac{\sqrt{\lambda}}{2\pi} \int d\sigma. \label{eqJ}
\eea
They satisfy a relation
\beq \label{esj}
E - J = \frac{S}{\omega} +K,
\eeq
where $K$ is a $\beta$-dependent correction term and given by the expression
\beq \label{eqK}
K = -\frac{\sqrt{\lambda}}{2\pi (1-v^2)} \int d\sigma \left\{
 \beta^2-
 \left(\beta \omega + \frac{\beta}{\omega}\right) \left(1-\frac{v^2}{1-\beta\omega}\right)
 \right\}.
\eeq
$K$ vanishes identically as $\beta=0$, and (\ref{esj}) reduces to $E - J = \frac{S}{\omega}$, consistent with \cite{Ryang:2006yq}.

\begin{figure}
\begin{center}
 \subfigure[ ]
  {
    \includegraphics[width=0.4\textwidth]{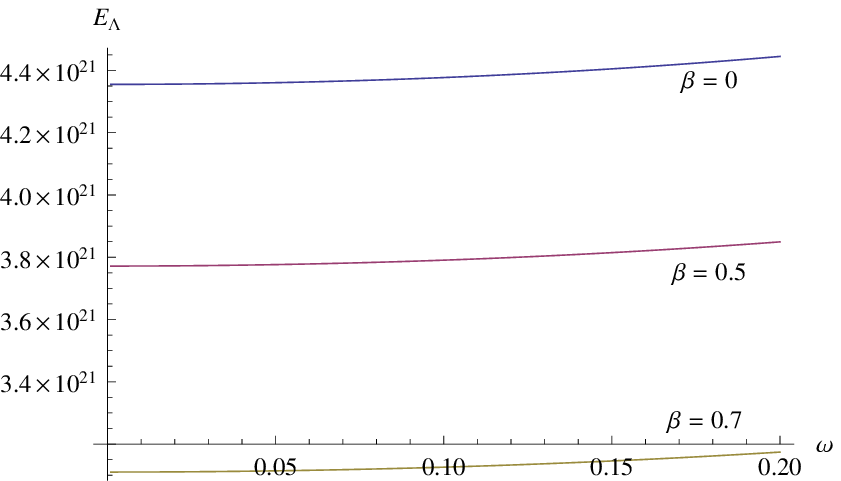}
    \label{fig:E1}
  }\qquad
  \subfigure[ ]
  {
     \includegraphics[width=0.4\textwidth]{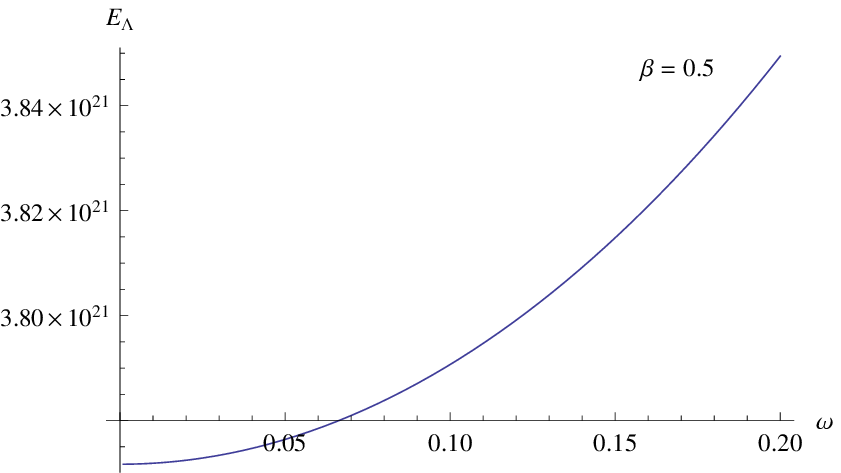}
     \label{fig:E2}
  }
\caption{$E_{\Lambda}$ of the spiky strings against $\omega$, with (a) three values of $\beta$, and with (b) $\beta=0.5$ magnified from (a). We've chosen $v=0.4$, and the integration cutoff $\Lambda$ is at $\rho=50$. $\frac{S_{\Lambda}}{\omega}$ behaves similarly against $\omega$ for sufficiently large $\omega$.
}
\label{fig:E}
\end{center}
\end{figure}


\begin{figure}
\begin{center}
 \subfigure[ ]
  {
    \includegraphics[width=0.4\textwidth]{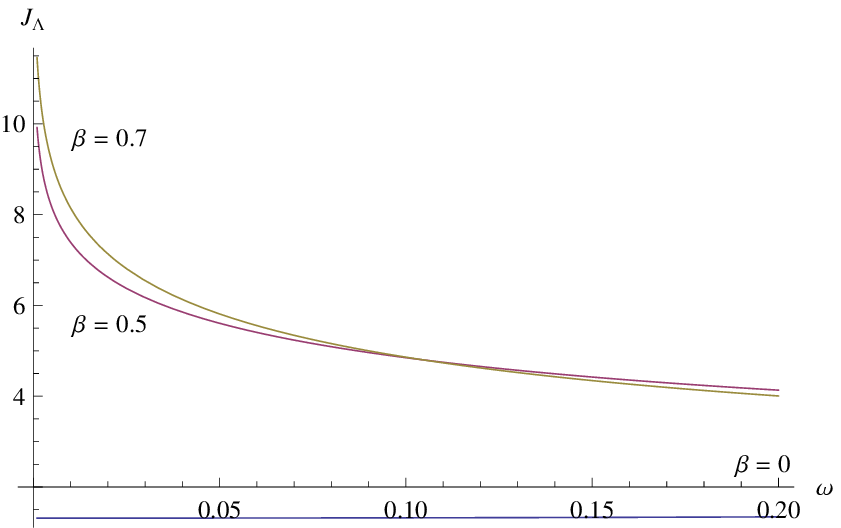}
    \label{fig:J1}
  }\qquad
  \subfigure[ ]
  {
     \includegraphics[width=0.4\textwidth]{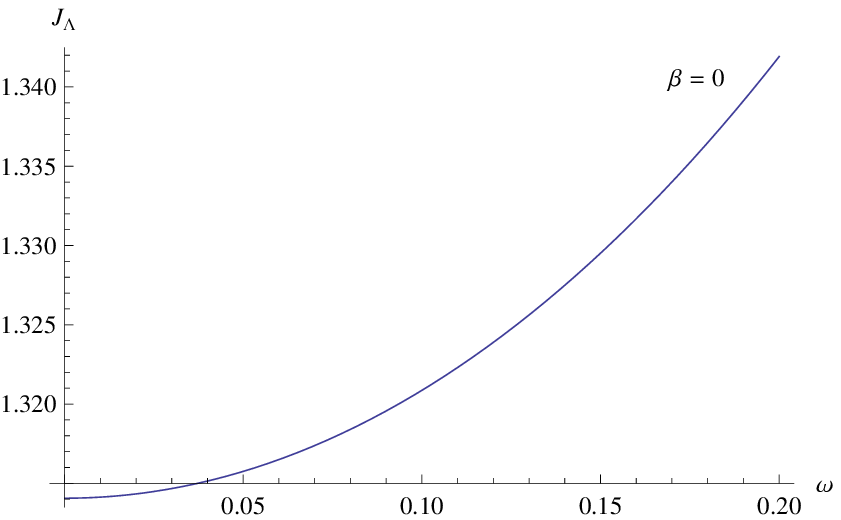}
     \label{fig:J0}
  }
\caption{$J_{\Lambda}$ of the spiky strings against $\omega$, with (a) three values of $\beta$, and with (b) $\beta=0$ magnified from (a). Here $v=0.4$, and $\Lambda$ is at $\rho=50$.
}
\label{fig:J}
\end{center}
\end{figure}

\begin{figure}
\begin{center}
     \includegraphics[width=0.4\textwidth]{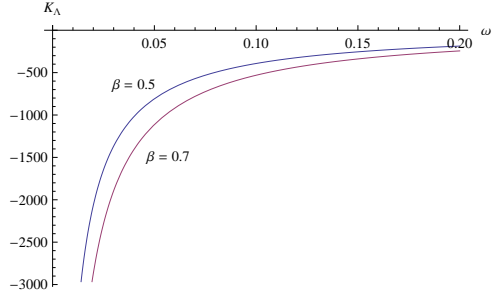}
\caption{$K_{\Lambda}$ of the spiky strings against $\omega$, with $\beta=0.5$ and $\beta=0.7$, at $v=0.4$. The integration cutoff is at $\rho=50$. The magnitude of $K_{\Lambda}$ increases as $\omega$ decreases, but $K_{\Lambda}=0$ identically at $\beta=0$.
}
\label{fig:K}
\end{center}
\end{figure}

(\ref{eqE})$\sim$(\ref{eqJ}) diverge as they are integrated to $\rho \to \infty$, and require regularization. However, due to lack of analytic results out of these integrals, we are unable to obtain analytic regularized expressions for $E$, $S$ and $J$. We refer to the numerical computation to reveal their dependence on $\omega$, such that the dispersion relation (\ref{esj}) is satisfied. We take a cutoff at $\Lambda = 50$ while integrating these quantities over $\rho$, and denote them by $E_{\Lambda}, S_{\Lambda}, J_{\Lambda},$ and $K_{\Lambda}$. The behaviors of $E$, $S$, $J$ against $\omega$ remains the same whether they are regularized or not, up to a large constant to be subtracted in regularization. The results are given in Figures \ref{fig:E} $\sim$ \ref{fig:K}.

\section{Analytical continuation and complex sin(h)-Gordon model}
While a string prapogates in metric (\ref{globalmetric}) is expected to have an equivalent description in terms of sinh-Gordon model, it is entertaining to see its connection to sine-Gordon model via analytic continuation.  With the coordinates change:
\begin{equation}\label{analytic_continue}
\rho \to i \gamma,\quad \phi \to \hat{t},\quad t \to \varphi_1,\quad \theta \to \varphi_2,
\end{equation}
One obtains the new metric
\begin{eqnarray}
ds^2 \to -\hat{ds}^2 = &&-d\hat{t}^2 + d\gamma^2 + [(1-\beta^2)\cos^2{\gamma} +\beta^2] d\varphi_1^2 \nonumber\\
&&+[(1-\beta^2)\sin^2{\gamma} + \beta^2]d\varphi_2^2 - 2\beta d\varphi_1 d\varphi_2,
\end{eqnarray}
whose spatial part is a squashed three-sphere\footnote{A different way to deform $S^3$ is by squashing the Hopf fiber $S^1$ along the $S^2$ base.  A different spin chain model could be also obtained\cite{Wen:2006fw}.}.  The deform parameter $\beta$ changes the geometry to a round sphere at $\beta=0$ and to a $3$-tori at $\beta=1$.   The spin chain at fast spinning limit after analytic continuation takes the following form:
\begin{eqnarray}
&&S=\frac{\sqrt{\lambda}}{4\pi} \int{d\tau d\sigma} 2 (1-\beta)\kappa \dot{\phi_1} + 2 (1+\beta) \cos{\gamma} \kappa \dot{\phi_2} \nonumber\\
&&- \frac{1}{4} \{  \gamma^{\prime 2} + (1-\beta)^2 \phi_1^{\prime 2} + (1+\beta)^2 \phi_2^{\prime 2} + 2(1-\beta^2)\cos{\gamma} \phi_1^\prime \phi_2^\prime \}
\end{eqnarray}
for another coordinates change $\phi_1 \to \varphi_1+\varphi_2,\quad \phi_2 \to \varphi_1-\varphi_2,\quad \gamma \to \gamma/2$ and boosting $\phi_1 \to \phi_1 + {\hat t}/(1-\beta)$.  Utilizing the Virasoro constraint, one obtains a deformed spin chain Hamiltonian density:
\begin{equation}
{\cal H} = (\partial_\sigma \gamma )^2 + (1+\beta)^2 \sin^2{\gamma} (\partial_\sigma \phi_2)^2,
\end{equation}
which recovers the $SU(2)$ Heisenberg XXX spin chain in \cite{Kruczenski:2003gt} for $\beta=0$.   One might wonder what is this deformed spin chain on the dual field theory side.  If one embeds the squashed sphere in $R^4$ as
\begin{eqnarray}
&&X^1 = \cos{\gamma}\cos{(\varphi_1-\beta \varphi_2)}, \quad X^2 = \sin{\gamma}\cos{(\varphi_2-\beta \varphi_1)},\nonumber\\
&&X^3 = \sin{\gamma}\sin{(\varphi_2-\beta \varphi_1)}, \quad X^4 = \cos{\gamma}\sin{(\varphi_1-\beta \varphi_2)}.
\end{eqnarray}
An academic guess is that the deformed spin chain is still made of the single trace operator $Tr(ZZZ\cdots)$ but with a twisted complex scalar $Z\equiv X^1 + i X^4 = \cos(\gamma)e^{i(\varphi_1-\beta\varphi_2)}$.

At last, the complex sine-Gordon model can be obtained by first constructing a new vector $K^i=\epsilon_{ijkl}X^j\partial_+ X^k \partial_- X^l$ out of $X^i$ and use them to define two $O(4)$-invariants $\phi$ and $\chi$:
\begin{equation}
\cos{\phi}\equiv -\partial_+\vec{X} \cdot \partial_-\vec{X}, \quad \pm 2 \partial_{\pm} \chi \sin^2{(\phi/2)} \equiv \partial^2_{\pm} \vec{X}\cdot \vec{K}.
\end{equation}
It can be shown that the equations of motion of $\phi$ and $\chi$ are nothing but the complex sine-Gordon equations\cite{Okamura:2006zv}.  By analytically continuing back to deformed $AdS_3\times S^1$, one can also obtain the sinh-Gordon equation as shown in \cite{Ryang:2006yq}.

\section{PP-wave limit}
In this section, we consider a spiky string solution on a pp-wave limit of the $\beta$-deformed $ AdS_3$ background, in which we especially see the region of $\rho \to \infty$ and $\theta \to t$. Following \cite{Losi:2010hr}, we start with the $\beta$-$ AdS_3$ part in the metric \eqref{globalmetric}, and consider the following reparametrization,
\begin{eqnarray}
z &=& \frac{ 2\sqrt{2} e^{\rho_0} }{ e^\rho }, \\
x_{\pm} &=& e^{\rho_0} e^{\mp \theta_0 } ( \theta \pm t ).
\end{eqnarray}
We take limits of the two parameters as $\rho_0, \theta_0 \to \infty$ with fixing the ratio
\begin{equation}
\frac{e^{\theta_0}}{e^{\rho_0}} = 2 \mu.
\end{equation}
Here $\mu > 0$ can be regarded as a free parameter. At this limit, we obtain $\beta$-AdS pp-wave metric
\begin{equation}
ds^2 = \frac{1}{z^2} \left\{ 2 ( 1- \beta^2 ) dx_+ dx_- + dz^2 - ( 1- \beta )^2 \mu^2 z^2 dx_+^2   \right\}.
\end{equation}
The boundary is located at $z=0$.

We shall consider a spiky string solution in the $\beta$-$ AdS_3$-pp-wave with the following ansatz:
\begin{equation}
x_+ =  \tau, \
x_- =  \sigma, \
z = z(\sigma), \
- \sigma_0 \leq \sigma \leq \sigma_0.
\end{equation}
The Nambu-Goto action in this case is given by
\begin{equation}
S = - \frac{\sqrt{\lambda}}{2 \pi} \int \frac{1}{z^2}
\sqrt{  ( 1- \beta )^2 z^2 \mu^2 ( \partial_\sigma z )^2 + ( 1 - \beta^2 )^2 }
d\tau d\sigma.
\end{equation}
From the equations of motion, we obtain the differential equation
\begin{equation}
\partial_{\sigma} z = ( 1+ \beta ) \frac{1}{z} \sqrt{ \frac{z_0^4}{z^4} -1  },
\end{equation}
and this can be solved by
\begin{equation}
z( \sigma ) = \sqrt{2} \sqrt{ 1+\beta } ( \sigma_0^2 - \sigma^2 )^{1/4}.
\end{equation}
We can evaluate conserved quantities associated with $x_+$ and $x_-$ as
\begin{eqnarray}
P_+
&=& - \frac{\sqrt{\lambda}}{2\pi } \int_{-\sigma_0}^{\sigma_0} \frac{ \sqrt{ ( 1 - \beta )^2 z^2 ( \partial_\sigma z )^2 + ( 1- \beta^2 )^2 }}{ z^2 } d \sigma \\
&=& - 2 \frac{\sqrt{\lambda}}{2\pi } ( 1 - \beta ) \int_\epsilon^{z_0} \frac{1}{ z \sqrt{ 1 - \frac{z^4}{z_0^4} } } dz, \\
P_-
&=&  \frac{\sqrt{\lambda}}{2\pi } ( 1- \beta  ) \int_{-\sigma_0}^{\sigma_0}
\frac{1}{ z^2 } \frac{ ( \partial_\sigma z )^2 }{ \sqrt{ ( 1-\beta )^2 z^2 ( \partial_\sigma z )^2 + ( 1- \beta^2)^2 }} d \sigma \\
&=&  2 \frac{\sqrt{\lambda}}{2\pi } \int_\epsilon^{z_0} \frac{1}{ z^3  } \sqrt{ 1 - \frac{z^4}{z_0^4} }  dz.
\end{eqnarray}
Here we set $\mu = 1$ and $z_0 = \sqrt{ 2 ( 1 + \beta ) \sigma_0 }$, and we also introduced a cutoff $\epsilon$ that should be taken $\epsilon \to 0$. By expanding by $\epsilon$ we get
\begin{eqnarray}
P_+
&\sim& - \frac{\sqrt{\lambda}}{2 \pi } ( 1 - \beta ) \ln \frac{z_0^2}{\epsilon^2} + \cdots, \\
P_-
&\sim&  \frac{\sqrt{\lambda}}{2 \pi }  \frac{ 1 }{\epsilon^2} - \frac{\sqrt{\lambda}}{2 \pi } \frac{\pi}{2z_0^2}  \cdots
.
\end{eqnarray}
From these expansions, we obtain the following relation
\begin{equation}
P_+ \sim - \frac{\sqrt{\lambda}}{2\pi } ( 1- \beta^2 ) \ln P_- + \cdots .
\end{equation}
This relation can be regarded as a $\beta$-deformed result from that in \cite{Kruczenski:2008bs}.

\begin{acknowledgments}
SHD wishes to thank Chong-Sun Chu, Shoichi Kawamoto and Jackson Wu for helpful discussions.  WYW is grateful to the hospitality and useful discussion with members in the CQUeST and YITP.   This project was partially supported by the Taiwan's National Science Council under Grant No. 102-2112-M-033-003-MY4 and National Center for Theoretical Sciences.

\end{acknowledgments}


\appendix*

\section{Equations of Motion and Virasoro Constraints from the Polyakov action}

In this appendix, we present the general expressions of the equations of motion and Virasoro constraint from the Polyakov action in the background (\ref{globalmetric}).

The Polyakov action is given by
\bea
I &=& - \frac{\sqrt{\lambda}}{4\pi} \int d\tau\, d\sigma\,
\left\{- [(1-\beta^2)\cosh^2\rho +\beta^2](t'{}^2-\dot{t}^2)
+ 2\beta(t'\theta' -\dot{t}\dot{\theta}) \right. \nn \\
&& \hspace{2.5cm} \left.+[(1-\beta^2)\sinh^2\rho - \beta^2] (\theta'{}^2 - \dot{\theta}^2)
+(\rho'{}^2 - \dot{\rho}^2) + (\phi'{}^2 - \dot{\phi}^2)\right\}, \label{action}
\eea
from which the equations of motion can be derived as follows:
\bea
\frac{\partial}{\partial \tau}\left[g_{tt} \frac{\partial t}{\partial \tau}\right] - \frac{\partial }{\partial \sigma}\left[ g_{tt} \frac{\partial t}{\partial \sigma} \right] + \beta\left[ \frac{\partial^2 \theta}{\partial \tau^2} -\frac{\partial^2 \theta}{\partial \sigma^2} \right]&=& 0, \label{eomt} \\
\frac{\partial }{\partial \tau}\left[g_{\theta\theta} \frac{\partial \theta}{\partial \tau}\right] - \frac{\partial }{\partial \sigma}\left[ g_{\theta\theta} \frac{\partial \theta}{\partial \sigma} \right] + \beta\left[ \frac{\partial^2 t}{\partial \tau^2}-\frac{\partial^2 t}{\partial \sigma^2} \right]&=& 0, \label{eomtheta} \\
(1-\beta^2)\,\sinh \rho \,\cosh \rho \,
\left[ ( \frac{\partial t}{\partial \tau} )^2 - ( \frac{\partial t}{\partial \sigma} )^2 - ( \frac{\partial \theta}{\partial \tau} )^2 +
( \frac{\partial \theta}{\partial \sigma} )^2 \right]
+\left(\frac{\partial^2 \rho}{\partial \tau^2}-\frac{\partial^2 \rho}{\partial \sigma^2}\right)&=&0, \label{eomrho} \\
\frac{\partial^2 \phi}{\partial \tau^2}-\frac{\partial^2 \phi}{\partial \sigma^2} &=& 0. \label{eomphi}
\eea
The Virasoro constraints read
\bea
 g_{tt} \frac{\partial t}{\partial \tau} \frac{\partial t}{\partial \sigma} + \beta \left(
 \frac{\partial t}{\partial \tau} \frac{\partial \theta}{\partial \sigma} + \frac{\partial \theta}{\partial \tau} \frac{\partial t}{\partial \sigma} \right) + g_{\theta\theta} \frac{\partial \theta}{\partial \tau} \frac{\partial \theta}{\partial \sigma}
 +\frac{\partial \rho}{\partial \tau} \frac{\partial \rho}{\partial \sigma} + \frac{\partial \phi}{\partial \tau} \frac{\partial \phi}{\partial \sigma}
 &=& 0, \label{v1} \\
 g_{tt} \left[(\frac{\partial t}{\partial \tau})^2+ (\frac{\partial t}{\partial \sigma})^2\right]
 + 2\beta \left[
 \frac{\partial t}{\partial \tau} \frac{\partial \theta}{\partial \tau} +  \frac{\partial t}{\partial \sigma} \frac{\partial \theta}{\partial \sigma} \right]
 +g_{\theta\theta} \left[(\frac{\partial \theta}{\partial \tau})^2+ (\frac{\partial \theta}{\partial \sigma})^2\right]&&\nn\\
 +(\frac{\partial \rho}{\partial \tau})^2+ (\frac{\partial \rho}{\partial \sigma})^2
 +(\frac{\partial \phi}{\partial \tau})^2+ (\frac{\partial \phi}{\partial \sigma})^2&=&0. \label{v2}
\eea


\end{document}